# Improved MST3 Encryption scheme based on small Ree groups

Gennady Khalimov[1[0000-0002-2054-9186]], Yevgen Kotukh [2[0000-0003-4997-620X]]

[1]Kharkiv National University of Radioelectronics, Kharkiv, 61166, Ukraine hennadii.khalimov@nure.ua
[2]Yevhenii Bereznyak Military Academy, Kyiv, Ukraine yevgenkotukh@gmail.com

**Abstract.** This article presents an encryption scheme based on the small Ree groups. We propose utilizing the small Ree group structure to enhance the overall security parameters of the encryption scheme. By extending the logarithmic signature to encompass the entire group and modifying the encryption algorithm, we have developed robust protection against sequential key recovery attacks.

## INTRODUCTION

Currently, most public cryptographic primitives rely on the presumed intractability of specific mathematical problems in very large finite abelian groups. These hard problems include factoring large integers, solving discrete logarithm problems over finite fields or elliptic curves, and similar challenges. However, with the development of quantum algorithms capable of factoring large integers and solving discrete logarithm problems, most established public-key cryptosystems will become vulnerable once practical quantum computers are implemented.

Since the 1980s, researchers have been developing cryptographic schemes based on difficult problems in group theory. In [1], an approach to design public-key cryptosystems based on groups and semigroups with undecidable word problems was introduced. This approach was further explored and enhanced in [2-6]. The algebraic properties of logarithmic signatures and related cryptosystems were specifically examined in [7,8]. In 2002, Magliveras et al. [9] introduced the public key cryptosystems MST1 and MST2. Subsequently, Lempken et al. [10] utilized logarithmic signatures and random covers to construct the generic MST3 encryption scheme.

MST cryptosystems have not demonstrated vulnerability to quantum algorithm attacks, positioning them as viable candidates for post-quantum public-key cryptography. A comprehensive analysis of this resistance is presented in [13]. The classical MST3 cryptosystem employs logarithmic signature computations in the center of the Suzuki group, which has a relatively large center. Implementations of MST3 cryptosystems using groups of automorphisms of the Suzuki, Hermitian, and Ree functional fields, as well as generalized Suzuki groups, are proposed in [21-23]. These works demonstrate that efficient encryption schemes with high security can be constructed using logarithmic signatures on multi-parameter groups of large orders.

The small Ree group offers three parametric representations and maintains a smaller center relative to its group order. In this paper, we present an encryption scheme based on small Ree groups of large order, utilizing logarithmic signatures outside the group's center.

## SMALL REE GROUPS

The small Ree groups were first described by Ree [16,17], with their basic construction detailed in [18]. The matrix generators for these groups are explicitly characterized in [17]. Let $q = 3^{2m+1}$ for some $m > 0$ and $t = 3^m$.

The small Ree group is defined as $Ree(q) = \langle \alpha(x), \beta(x), \gamma(x), h(\lambda), I^- | x \in F_q, \lambda \in F_q^\times \rangle$.

Subgroup U(q) for the group Ree(q) of upper triangular matrices $\alpha(x), \beta(x), \gamma(x)$ has presentation $U(q) = \langle \alpha(x), \beta(x), \gamma(x) | x \in F_q \rangle$.

Each element of $U(q)$ can be expressed uniquely $S(a, b, c) = \alpha(a)\beta(b)\gamma(c)$ so $U(q) = \{S(a, b, c) | a, b, c \in F_q\}$, and it follows that $|U(q)| = q^3$. Also, $U(q)$ is a Sylow 3–subgroup of $Ree(q)$, and direct calculations show that

$S(a_1, b_1, c_1)S(a_2, b_2, c_2) = S(a_1 + a_2, b_1 + b_2 - a_1 a_2^{3t}, c_1 + c_2 - a_2 b_1 + a_1 a_2^{3t+1} - a_1^2 a_2^{3t}), S(a, b, c)^{-1} = S(-a, -b - a^{3t+1}, -c - ab + a^{3t+2})$.

The center $Z(U(q)) = \{S(0,0,c) | c \in F_q\}$.

The deriver group $U_1(q) = \{S(0,b,c) | b, c \in F_q\}$, it follows that $|U_1(q)| = q^2$ and its elements have order 3. The elements in $U(q) \setminus U_1(q) = \{S(a,b,c) | a \neq 0\}$ have order 9.

The subgroup $U(q)$ for the small group $Ree(q)$ has a greater $ord U(q) = q^3$ then the orders of the Suzuki groups and the automorphism group $A(P_\infty)$ of the Hermitian function field $H|F_{q^2}$. Suzuki groups, which appear in MST3 cryptosystems, are isomorphic to the projective linear group $PGL(3, F_q)$, where $q = 2q_0^2$, $q_0 = 2^n$ and has order $q^2$. The automorphism group of the Hermitian function field $H|F_{q^2}$, has a $ord A(P_\infty) = q^3(q^2 - 1)$. Several encryption algorithms have been proposed for MST cryptosystems, with the latest version, known as MST3, having been implemented for the Suzuki group [6]. This paper proposes an alternative encryption scheme based on small Ree groups. Our approach utilizes logarithmic signatures for encryption across coordinates outside the group's center. The subsequent chapter presents a detailed discussion of our proposed solution, including practical computational aspects.

## PROPOSED SOLUTION

We will show the correctness of the obtained expressions in the following simple example. Fix the subgroup $U(q) = \{S(a,b,c) | a,b,c \in F_q\}$ for the group $Ree(q)$ over $F_q$, $q = 3^5$, $g(x) = x^5 + 2x + 1$, $t = 3^2$. Group operation is defined as a product of two matrices $S(a_1,b_1,c_1)S(a_2,b_2,c_2) = S(a_1 + a_2, b_1 + b_2 - a_1 a_2^{3t}, c_1 + c_2 - a_2 b_1 + a_1 a_2^{3t+1} - a_1^2 a_2^{3t})$.

The inverse element is determined as $S(a,b,c)^{-1} = S(-a, -b - a^{3t+1}, -c - ab + a^{3t+2})$.

The identity is the triple $S(0,0,0)$.

Step 1. We construct tame logarithmic signatures $\beta_{(k)} = [B_{1(k)},...,B_{s(k)}] = (b_{ij})_{(k)}$, $b_{ij(k)} \in F_q$, $i = \overline{1, s(k)}$, $j = \overline{1, r_{i(k)}}$, $k = \overline{1,2}$ of types $(r_{1(k)},...,r_{s(k)})$ for coordinate $b$, $c$.

Logarithmic signatures $\beta_1$ and $\beta_2$ in a group representations define $b_{ij(1)}$ and $b_{ij(2)}$ coordinates. Types $(r_{1(k)},...,r_{s(k)})$ and logarithmic signatures $\beta_1$ and $\beta_2$ are chosen independently. Let`s logarithmic signatures $\beta_1$ and $\beta_2$, as an example, have a types $(r_{1(1)}, r_{2(1)}, r_{3(1)}) = (3^2, 3^2, 3)$, $(r_{1(2)}, r_{2(2)}, r_{3(2)}) = (3, 3^2, 3^2)$. Arrays $b_{ij(1)}$ and $b_{ij(2)}$ consists of three subarrays with a number of rows equal to $r_i$. You can select any fragmentation of arrays with the condition $\prod_{i=1}^{s} r_i = q$. In our case we have $\prod_{i=1}^{s} r_i = 3^5$. Each row $b_{ij}$ it`s an element of the field $F_q$. The construction of arrays of logarithmic signatures is presented in [19]. First stage is to generate a tame logarithmic signature with the dimension of corresponding selected type $(r_{1(k)},...,r_{s(k)})$ and finite field $F_q$. To increase the security of arrays $\beta_k$ various cryptographic transformations can be used. For example, simple ones like adding noise vectors, permutations of strings in subarrays $B_{i(j)}$, merge of arrays $B_{i(j)}$, their permutation, matrix transformations. In our example, we use array noise. This allows you to build two different logarithmic signatures $\beta_k = [B_{1(k)}, B_{2(k)}, B_{3(k)}]$. Arrays of logarithmic signatures $\beta_k$ in the group representation, defines the coordinates $b_{ij(k)}$, respectively $\beta_1 = [B_{1(1)},...,B_{s(1)}] = (b_{ij})_{(1)} = S(0, b_{ij(1)}, 0)$

$$\beta_2 = [B_{1(2)},...,B_{s(2)}] = (b_{ij})_{(2)} = S(0, 0, b_{ij(2)}).$$

| $\beta_1=$ | $b_{ij(1)}$ | | | $S(0,b_{ij(1)},0)$ | | $\beta_2=$ | $b_{ij(2)}$ | | | $S(0,0,b_{ij(2)})$ |
|---|---|---|---|---|---|---|---|---|---|---|
| B1(1) | 00 | 00 | 0 | 0,0,0 | | B1(2) | 0 | 00 | 00 | 0,0,0 |
| | 10 | 00 | 0 | 0,$\alpha^0$,0 | | | 1 | 00 | 00 | 0,0,$\alpha^0$ |
| | 20 | 00 | 0 | 0,$\alpha^{121}$,0 | | | 2 | 00 | 00 | 0,0,$\alpha^{121}$ |
| | 01 | 00 | 0 | 0,$\alpha^1$,0 | | B2(2) | 0 | 00 | 00 | 0,0,0 |
| | 11 | 00 | 0 | 0,$\alpha^{69}$,0 | | | 2 | 10 | 00 | 0,0,$\alpha^5$ |
| | 21 | 00 | 0 | 0,$\alpha^5$,0 | | | 2 | 20 | 00 | 0,0,$\alpha^{190}$ |
| | 02 | 00 | 0 | 0,$\alpha^{122}$,0 | | | 1 | 01 | 00 | 0,0,$\alpha^{46}$ |
| | 12 | 00 | 0 | 0,$\alpha^{126}$,0 | | | 0 | 11 | 00 | 0,0,$\alpha^{70}$ |
| | 22 | 00 | 0 | 0,$\alpha^{190}$,0 | | | 2 | 21 | 00 | 0,0,$\alpha^{222}$ |
| B2(1) | 21 | 00 | 0 | 0,$\alpha^5$,0 | | | 1 | 02 | 00 | 0,0,$\alpha^{195}$ |
| | 12 | 10 | 0 | 0,$\alpha^{138}$,0 | | | 2 | 12 | 00 | 0,0,$\alpha^{17}$ |
| | 02 | 20 | 0 | 0,$\alpha^{191}$,0 | | | 2 | 22 | 00 | 0,0,$\alpha^{131}$ |
| | 12 | 01 | 0 | 0,$\alpha^{198}$,0 | | B3(2) | 2 | 12 | 00 | 0,0,$\alpha^{17}$ |
| | 01 | 11 | 0 | 0,$\alpha^{11}$,0 | | | 1 | 21 | 10 | 0,0,$\alpha^{30}$ |
| | 20 | 21 | 0 | 0,$\alpha^{36}$,0 | | | 1 | 02 | 20 | 0,0,$\alpha^{109}$ |
| | 20 | 02 | 0 | 0,$\alpha^{86}$,0 | | | 2 | 22 | 01 | 0,0,$\alpha^{105}$ |
| | 11 | 12 | 0 | 0,$\alpha^{39}$,0 | | | 0 | 10 | 11 | 0,0,$\alpha^{228}$ |
| | 11 | 22 | 0 | 0,$\alpha^{22}$,0 | | | 1 | 02 | 21 | 0,0,$\alpha^{154}$ |
| B3(1) | 01 | 12 | 0 | 0,$\alpha^{102}$,0 | | | 1 | 01 | 02 | 0,0,$\alpha^{206}$ |
| | 02 | 20 | 1 | 0,$\alpha^{150}$,0 | | | 2 | 00 | 12 | 0,0,$\alpha^{220}$ |
| | 22 | 20 | 2 | 0,$\alpha^{21}$,0 | | | 1 | 02 | 22 | 0,0,$\alpha^{239}$ |

Step 2. We construct random covers $\alpha_k$, for the same type as $\beta_1$ и $\beta_2$

$$\alpha_1 = [A_{1(1)},...,A_{s(1)}] = (a_{ij})_{(1)} = S(a_{ij(1)_1}, a_{ij(1)_2}, a_{ij(1)_3}), \quad \alpha_2 = [A_{1(2)},...,A_{s(2)}] = (a_{ij})_{(2)} = S(0, a_{ij(2)_2}, a_{ij(2)_3}),$$

where $a_{ij} \in U(q)$, $a_{ij(k)_1}, a_{ij(k)_2}, a_{ij(k)_3} \in F_q \setminus \{0\}$, $i = \overline{1,s}$, $j = \overline{1, r_{(k)}}$, $k = \overline{1,2}$.

Each cover $\alpha_k$ defined by three arrays $(a_{ij(k)_1}, a_{ij(k)_2}, a_{ij(k)_3})$ with non-zero entries.

Let`s generate random covers $\alpha_1 = [A_{1(1)}, A_{2(1)}, A_{3(1)}]$, $\alpha_2 = [A_{1(2)}, A_{2(2)}, A_{3(2)}]$.

In the field representation $\alpha_k = S(a_{ij(k)_1}, a_{ij(k)_2}, a_{ij(k)_3})$, $k = \overline{1,2}$ has the following form

| $\alpha_1 = [A_{1(1)}, A_{2(1)}, A_{3(1)}]$ | | | $\alpha_2 = [A_{1(2)}, A_{2(2)}, A_{3(2)}]$ | | |
|---|---|---|---|---|---|
| A1(1) | A2(1) | A3(1) | A1(2) | A2(2) | A3(2) |
| $\alpha^{48}, 0, \alpha^{26}$ | $\alpha^{165}, \alpha^{28}, \alpha^{21}$ | $\alpha^{78}, \alpha^{205}, \alpha^{15}$ | $0, \alpha^{139}, \alpha^{205}$ | $0, \alpha^{131}, \alpha^{132}$ | $0, \alpha^{241}, \alpha^{96}$ |
| $\alpha 61, \alpha 11, \alpha 159$ | $\alpha^{204}, \alpha^{176}, \alpha^{228}$ | $\alpha^1, \alpha^{26}, \alpha^{114}$ | $0, \alpha^{106}, \alpha^{210}$ | $0, \alpha^{133}, \alpha^{177}$ | $0, \alpha^{197}, \alpha^{165}$ |
| $\alpha^{233}, \alpha^{206}, \alpha^{67}$ | $\alpha^{135}, \alpha^{126}, \alpha^{115}$ | $\alpha^{166}, \alpha^{38}, \alpha^{31}$ | $0, \alpha^{86}, \alpha^{171}$ | $0, \alpha^{198}, \alpha^{96}$ | $0, \alpha^{117}, \alpha^{126}$ |
| $\alpha^{165}, \alpha^{204}, \alpha^{190}$ | $\alpha^{215}, \alpha^{208}, \alpha^{99}$ | | | $0, \alpha^{101}, \alpha^{165}$ | $0, \alpha^{155}, \alpha^{152}$ |
| $\alpha^6, \alpha^1, \alpha^{78}$ | $\alpha^{127}, \alpha^{69}, \alpha^{103}$ | | | $0, \alpha^{32}, \alpha^{88}$ | $0, \alpha^{156}, \alpha^{95}$ |
| $\alpha^{132}, \alpha^{85}, \alpha^{65}$ | $\alpha^{150}, \alpha^{80}, \alpha^{206}$ | | | $0, \alpha^{239}, \alpha^{11}$ | $0, \alpha^{216}, \alpha^{34}$ |
| $\alpha^{24}, \alpha^{12}, \alpha^{79}$ | $\alpha^{150}, \alpha^{43}, \alpha^{186}$ | | | $0, \alpha^{233}, \alpha^{85}$ | $0, \alpha^{24}, \alpha^{226}$ |
| $\alpha^{190}, \alpha^{211}, \alpha^{216}$ | $\alpha^{54}, \alpha^{61}, \alpha^{34}$ | | | $0, \alpha^0, \alpha^{230}$ | $0, \alpha^{240}, \alpha^{55}$ |
| $\alpha^{19}, \alpha^{104}, \alpha^{98}$ | $\alpha^7, \alpha^{51}, \alpha^{108}$ | | | $0, \alpha^{110}, \alpha^{93}$ | $0, \alpha^{35}, \alpha^{168}$ |

Step 3. Choose random $t_{0(k)}, t_{1(k)}, ..., t_{s(k)} \in U(q) \setminus Z$, $s = 3$, $k = \overline{1,2}$ and $t_{3(1)} = t_{0(2)}$.

For the first logarithmic signatures $\beta_1, \beta_2$ we have

| | | | |
|---|---|---|---|
| t0(1)=($\alpha^{123}, \alpha^{31}, \alpha^{51}$) | t-10(1)=($\alpha^2, \alpha^{218}, \alpha^{170}$) | t0(2)=($\alpha^{241}, \alpha^{69}, \alpha^{45}$) | t-10(2)=($\alpha^{120}, \alpha^{28}, \alpha^{35}$) |
| t1(1)=($\alpha^{133}, \alpha^{94}, \alpha^{26}$) | t-11(1)=($\alpha^{12}, \alpha^{94}, \alpha^{147}$) | t1(2)=($\alpha^{206}, \alpha^{130}, \alpha^{106}$) | t-11(2)=($\alpha^{85}, \alpha^{174}, \alpha^{19}$) |
| t2(1)=($\alpha^{205}, \alpha^{149}, \alpha^{164}$) | t-12(1)=($\alpha^{84}, \alpha^{94}, \alpha^{214}$) | t2(2)=($\alpha^{49}, \alpha^{10}, \alpha^{180}$) | t-12(2)=($\alpha^{170}, \alpha^{228}, \alpha^{211}$) |
| t3(1)=($\alpha^{241}, \alpha^{69}, \alpha^{45}$) | t-10(1)=($\alpha^{120}, \alpha^{28}, \alpha^{35}$) | t3(2)=($\alpha^{97}, \alpha^{43}, \alpha^{118}$) | t-13(2)=($\alpha^{218}, \alpha^{37}, \alpha^{113}$) |

Step 4. Calculate the arrays $\gamma_k$

$$\gamma_k = [h_{1(k)},...,h_{3(k)}] = (h_{ij})_k = t_{(i-1)(k)}^{-1} f((a_{ij})_k)(b_{ij})_k t_{i(k)}, \quad k = \overline{1,2}$$

Step 5. Construct a homomorphism $f$ defined by $f(S(a,b,c)) = S(0,a,b)$.

In the field representation $\gamma_1 = S(h_{ij(1)_1}, h_{ij(1)_2}, h_{ij(1)_3})$ and $\gamma_2 = S(h_{ij(2)_1}, h_{ij(2)_2}, h_{ij(2)_3})$ has the following form

| $\gamma_1 = S(h_{ij(1)_1}, h_{ij(1)_2}, h_{ij(1)_3})$ | | | $\gamma_2 = S(h_{ij(2)_1}, h_{ij(2)_2}, h_{ij(2)_3})$ | | |
|---|---|---|---|---|---|
| h1(1) | h2(1) | h3(1) | h1(2) | h2(2) | h3(2) |
| $\alpha^{193}, \alpha^{238}, \alpha^{29}$ | $\alpha^{10}, \alpha^{15}, \alpha^{83}$ | $\alpha^{75}, \alpha^5, \alpha^{168}$ | $\alpha^2, \alpha^{160}, \alpha^{106}$ | $\alpha^{56}, \alpha_{56}, \alpha^{209}$ | $\alpha^{63}, \alpha^{68}, \alpha^{185}$ |
| $\alpha^{193}, \alpha^4, \alpha^{96}$ | $\alpha^{10}, \alpha^{212}, \alpha^{82}$ | $\alpha^{75}, \alpha^{141}, \alpha^{135}$ | $\alpha^2, \alpha^{160}, \alpha^{131}$ | $\alpha^{56}, \alpha^{56}, \alpha^{146}$ | $\alpha^{63}, \alpha^{68}, \alpha^{169}$ |
| $\alpha^{193}, \alpha^{42}, \alpha^{166}$ | $\alpha^{10}, \alpha^{215}, \alpha^{43}$ | $\alpha^{75}, \alpha^{231}, \alpha^{57}$ | $\alpha^2, \alpha^{160}, \alpha^{122}$ | $\alpha^{56}, \alpha^{56}, \alpha^7$ | $\alpha^{63}, \alpha^{68}, \alpha^{26}$ |
| $\alpha^{193}, \alpha^{213}, \alpha^{134}$ | $\alpha^{10}, \alpha^{210}, \alpha^{185}$ | | | $\alpha^{56}, \alpha^{56}, \alpha^{167}$ | $\alpha^{63}, \alpha^{68}, \alpha^{223}$ |
| $\alpha^{193}, \alpha^{203}, \alpha^{19}$ | $\alpha^{10}, \alpha^{141}, \alpha^{81}$ | | | $\alpha^{56}, \alpha^{56}, \alpha^{32}$ | $\alpha^{63}, \alpha^{68}, \alpha^{123}$ |
| $\alpha^{193}, \alpha^{231}, \alpha^{180}$ | $\alpha^{10}, \alpha^{115}, \alpha^{162}$ | | | $\alpha^{56}, \alpha^{56}, \alpha^{96}$ | $\alpha^{63}, \alpha^{68}, \alpha^{26}$ |
| $\alpha^{193}, \alpha^{167}, \alpha^{214}$ | $\alpha^{10}, \alpha^{22}, \alpha^{144}$ | | | $\alpha^{56}, \alpha^{56}, \alpha^{132}$ | $\alpha^{63}, \alpha^{68}, \alpha^{92}$ |
| $\alpha^{193}, \alpha^{179}, \alpha^{133}$ | $\alpha^{10}, \alpha^{61}, \alpha^{232}$ | | | $\alpha^{56}, \alpha^{56}, \alpha^2$ | $\alpha^{63}, \alpha^{68}, \alpha^{212}$ |
| $\alpha^{193}, \alpha^1, \alpha^{70}$ | $\alpha^{10}, \alpha^{197}, \alpha^{209}$ | | | $\alpha^{56}, \alpha^{56}, \alpha^{177}$ | $\alpha^{63}, \alpha^{68}, \alpha^{15}$ |

For example, let $R_1 = 29$. We obtain the following basis factorization for a given type $(r_{1(1)}, r_{2(1)}, r_{3(1)}) = (3^2, 3^2, 3)$ in the form of $R_1 = (R_{1(1)}, R_{2(1)}, R_{3(1)}) = (2,3,0)$, where $R_1 + R_2 3^2 + R_3 3^4 = 29$. Compute $\gamma_1$

$\gamma_1(29) = h_{1(1)}(2) h_{2(1)}(3) h_{3(1)}(0) = S(\alpha^{193}, \alpha^{42}, \alpha^{166}) S(\alpha^{10}, \alpha^{210}, \alpha^{185}) S(\alpha^{75}, \alpha^5, \alpha^{168}) = S(\alpha^{206}, \alpha^{106}, \alpha^{219})$.

Let $R_2 = 31$. We obtain the $R_2 = (R_{1(2)}, R_{2(2)}, R_{3(2)}) = (1,1,1) = 31$ for a given type $(r_{1(2)}, r_{2(2)}, r_{3(2)}) = (3, 3^2, 3^2)$. Compute $\gamma_2$

$\gamma_2(31) = h_{1(2)}(1) h_{2(2)}(1) h_{3(2)}(1) = S(\alpha^{18}, \alpha^{154}, \alpha^{151})$.

For the encryption stage we have a following input: a message $m \in U_1(q)$, $m = S(0, m_2, m_3)$, and the public key $[f, (\alpha_k, \gamma_k)]$, $k = \overline{1,2}$. So, we are going to have ciphertext $(y_1, y_2)$ of the message $x$ as an output.

Let $m = (0, a^0, a^1) = S(0, a^0, a^1)$.

Choose a random $R = (R_1, R_2) = (29, 31)$, $R_1, R_2, \in \mathbb{Z}_{|F_q|}$.

Compute $y_1 = \alpha'(R) \cdot m = \alpha_1'(R_1) \cdot \alpha_2'(R_2) \cdot m = S(\alpha^{86}, \alpha^{186}, \alpha^{113})$,

$y_2 = \gamma'(R) = \gamma_1'(R_1) \cdot \gamma_2'(R_2) = S(\alpha^{238}, \alpha^{210}, \alpha^0)$, $y_3 = f(\alpha_2'(R_2)) = S(0, 0, \alpha^{66})$.

Output $y_1 = (\alpha^{86}, \alpha^{186}, \alpha^{113})$, $y_2 = (\alpha^{238}, \alpha^{210}, \alpha^0)$, $y_3 = (0, 0, \alpha^{66})$.

For the decryption stage we have a ciphertext $(y_1, y_2, y_3)$ and private key $[\beta_k, (t_{0(k)}, ..., t_{3(k)})]$, $k = \overline{1,2}$. So, we expect a message $m \in U_1(q)$ corresponding to ciphertext $(y_1, y_2, y_3)$.

To decrypt a message $m$, we need to restore random numbers $R = (R_1, R_2)$.

Compute $D^{(1)}(R_1, R_2) = t_{0(1)} y_2 t_{S(2)}^{-1} = t_{0(1)} S(\alpha^{238}, \alpha^{210}, \alpha^0) t_{S(2)}^{-1} = S(0, \alpha^{233}, \alpha^{143})$,

$$D^*(R) = f(y_1)^{-1} D^{(1)}(R_1, R_2) = S(0, \alpha^{86}, \alpha^{186}) S(0, \alpha^{233}, \alpha^{143})$$
$$= S(0, \alpha^2, \alpha^{176}).$$

We get $\beta_1(R_1) = \alpha^2 = (00100)$.

Recovery of $R_1$ was done earlier $R_1 = (R_{1(1)}, R_{2(1)}, R_{3(1)}) = (2,3,0)$.

| | |
|---|---|
| 00\|10\|0 | R1=(*,*,0) |
| 01\|12\|0 | row 0 from B3(1) |
| 00\|10\|0−01\|12\|0=02\|01\|0 | R1= (*,3,0) |
| 12\|01\|0 | row 3 from B2(1) |
| 02\|01\|0−12\|01\|0=20\|00\|0 | R1=(2,3,0) |

For further calculations, it is necessary to remove the components of the arrays $\alpha_1'(R_1)$ and $\gamma_1'(R_1)$ from ciphertext $(y_1, y_2)$.

Compute

$$y_2^{(1)} = \gamma_1'(R_1)^{-1}y_2 = S(\alpha^{206}, \alpha^{106}, \alpha^{219})^{-1}S(\alpha^{238}, \alpha^{210}, \alpha^0) =$$
$$S(\alpha^{85}, \alpha^{171}, \alpha^{11})S(\alpha^{238}, \alpha^{210}, \alpha^0) = S(\alpha^{18}, \alpha^{154}, \alpha^{151}).$$

Repeat the calculations

$$D^{(2)}(R_2) = t_{0(2)}y_2^{(1)}t_{s(2)}^{-1} = t_{0(2)}S(\alpha^{18}, \alpha^{154}, \alpha^{151})t_{s(l/2)}^{-1} = S(0,0,\alpha^8)$$

and

$$D^*(R) = D^{(2)}(R_2)y_3^{-1} = D^{(2)}(R_2)S(0,0, a_{(2)_2}(R_2))^{-1} =$$
$$S(0,0,\alpha^8)S(0,0,\alpha^{66})^{-1} = S(0,0,\alpha^8)S(0,0,\alpha^{187}) = S(0,0,\alpha^{227}).$$

Restore $R_2$ with $\beta_2(R_2) = \alpha^{227} = (10110)$.

Perform inverse calculations $\beta_2(R_2)^{-1}$. Select bit groups in vector $\beta(R)$ according to type $(r_{1(2)}, ..., r_{s(2)}) = (3, 3^2, 3^2)$. We use the same calculations as in the example for $\beta_1(R_1)^{-1}$, and we get

| | |
|---|---|
| 1\|01\|10 | R2=(*,*,1) |
| 1\|21\|10 | row 1 from B3(2) |
| 1\|01\|10−1\|21\|10=0\|10\|00 | R2= (*,1,1) |
| 2\|10\|00 | row 1 from B2(2) |
| 0\|10\|00−2\|10\|00=1\|00\|00 | R2=(1,1,1) |

$\beta_2(R')^{-1} = 1|01|10 = (R_{1(2)}, R_{2(2)}, R_{3(2)}) = (1,1,1)$.

$$R_2 = (R_{1(2)}, R_{2(2)}, R_{3(2)}) = (1,1,1) = 31.$$

Receive a message $m = \alpha'(R)^{-1}y_1 = \alpha_2'(R_2)^{-1} \cdot \alpha_1'(R_1)^{-1} \cdot y_1$

$$= S(0, \alpha^{66}, \alpha^{139})^{-1} S(\alpha^{86}, \alpha^{34}, \alpha^{217})^{-1} S(\alpha^{86}, \alpha^{186}, \alpha^{113}) = S(0, \alpha^0, \alpha^1).$$

So, message $m = (0, a^0, a^1)$ is expected output.

## SECURITY ANALYSIS

We consider the following types of attacks. Brute force attack on cipher text. By selecting $R = (R_1, R_2)$ try to decipher the text $y_1 = \alpha'(R) \cdot m = \alpha_1'(R_1) \cdot \alpha_2'(R_2) \cdot m$. For this case we consider complexity of attack implementation equal to $q^2$. Brute force attack on $R = (R_1, R_2)$. Select $R = (R_1, R_2)$ to match $y_2 = \gamma'(R) = \gamma_1'(R_1) \cdot \gamma_2'(R_2)$. For this case we also consider complexity of attack implementation equal to $q^2$. Third possible attack is to choose $R_1$ to match the value of $a_{(1)_1}(R_1)$ in the vector $y_1$ and choose $R_2$ to match the value of $a_{(2)_2}(R_2)$ in the vector $y_2$. For this case we consider complexity of attack implementation equal to $q$. But it can be improved with a possible link of $R_1$ and $R_2$ through matrix transformation. For this case, complexity of attack implementation equal $q^2$ also. Also, there is a way to brute force on $(t_{0(k)}, \ldots, t_{s(k)})$. For this case we consider complexity of attack implementation equal to $q^3$. Another possible attack is attack on the algorithm. Extraction parameters $a_{(1)_1}(R_1)$, $a_{(2)_2}(R_2)$ of $y_1$, $y_2$ does not allow to calculate $\alpha_1'(R_1) \cdot \alpha_2'(R_2)$ in $y_1 = \alpha_1'(R_1) \cdot \alpha_2'(R_2) \cdot m$. So, simple search of parameters $R_1, R_2$ leads to brute force attack with complexity $q^2$. Since the Ree group is defined over a large field $F_q$, the attack is not computationally possible.

## CONCLUSION

The encryption scheme implementation based on small Ree groups represents our contribution to the ongoing development of the established MST3 cryptosystem. We propose encrypting two logarithmic signatures that extend beyond the group's center, thereby increasing the encrypted message length and enhancing security. The length ciphertext is $2 \log q$ for computing in the finite field over $F_q$. The implementation of a cryptosystem on a group of Ree requires the construction of a logarithmic signature $\beta$ on vectors $3^h$, where $h$ is determined by the size of type $r_i = 3^h$. All blocks $B_i$ are subgroups of the $U_1(q) = \{S(0,b,c) | b,c \in F_q\}$. The size of the arrays $\beta$ and α is determined by the type $(r_1,...,r_s)_b$ and $(r_1,...,r_s)_c$ for coordinate $b,c$ for the subgroups $U_1(q)$. For 128-bit cryptography, which is equivalent to calculations over field $q = 3^{80}$, if the $r_i$ type is $r_i = 3^5 = 243$, $s = 8$, only the 1944 entries are required for cryptography on the group. A common disadvantage of cryptosystems on logarithmic signatures is the large size of key data.